# A combined experimental and theoretical study of the electronic and vibrational properties of bulk and few-layer Td-WTe$_2$

Manoj K. Jana[1], Anjali Singh[2], Dattatray J. Late[3], Catherine Rajamathi[4], Kanishka Biswas[1], Claudia Felser[4], Umesh V. Waghmare[2] and C. N. R. Rao[1*]

[1] New chemistry Unit, International Centre for Materials Science and Sheikh Saqr Laboratory, Jawaharlal Nehru Centre for Advanced Scientific Research (JNCASR), Bangalore 560 064, India.

[2] Theoretical Sciences Unit, Jawaharlal Nehru Centre for Advanced Scientific Research (JNCASR), Bangalore 560 064, India.

[3] Physical and Materials Chemistry Division, CSIR-National Chemical Laboratory, Pune 411008, India.

[4] Max Planck Institute of Chemical Physics for Solids, Dresden 01187, Germany

*Email: cnrrao@jncasr.ac.in

**Abstract**

The recent discovery of non-saturating giant positive magnetoresistance in Td-WTe$_2$ has aroused great interest in this material. We have studied the structural, electronic and vibrational properties of bulk and few-layer Td-WTe$_2$ experimentally and theoretically. Spin-orbit coupling is found to govern the semi-metallic character of Td-WTe$_2$. Its structural link with the metallic 1T form provides an understanding of its structural stability. We observe a metal to insulator transition and a change in the sign of the Seebeck coefficient around 373 K. Lattice vibrations in Td-WTe$_2$ have been analyzed by first principle calculations. Out of the 33 possible zone-center Raman active modes, five distinct Raman bands are observed around 112, 118, 134, 165 and 212 cm$^{-1}$ in bulk Td-WTe$_2$. Based on symmetry analysis and the calculated Raman tensors, we assign the intense bands at 165 cm$^{-1}$ and 212 cm$^{-1}$ to the $A_1^{'}$ and $A_1^{''}$ modes respectively. We have examined the effect of temperature and the number of layers on the Raman spectrum. Most of the bands of Td-WTe$_2$ stiffen, and the ratio of the integrated intensities of the $A_1^{''}$ to $A_1^{'}$ bands decreases in the few-layer sample, while all the bands soften in both bulk and few-layer samples with increasing temperature.

**Introduction**

Layered transition metal dichalcogenides (TMDCs) are inorganic analogues of graphene, with a wide range of electronic, optical, chemical, thermal and catalytic properties of fundamental and technological importance [1-4]. Among group VI dichalcogenides, the ditellurides exhibit eccentricity with respect to structure and properties and are relatively less studied to date. The early electronic structure calculations by Dawson and Bullet [5] revealed that unlike Group VI disulfides and diselenides, the ditellurides deviate from a simple band model predicting a semiconducting behavior due to trigonal prismatic crystal-field splitting. Tungsten ditelluride ($WTe_2$) crystallizes in a distorted variant of $CdI_2$-type structure with an octahedral coordination around the metal, referred to as Td-polytype [6]. The structure of Td-$WTe_2$ constitutes triple-layers of covalently bonded Te-W-Te atomic planes, stacked along c-axis through weak van der Waals interactions. The $WTe_6$ octahedra are strongly distorted due to off-centering of W atoms as the latter move towards each other to form slightly buckled W-W zigzag chains running along a-axis. Consequently, $WTe_2$ exhibits metallic bonding with a W-W bond distance of 2.849 Å - only about 0.13 Å longer than that in tungsten metal [7]. The reduced Madelung energy as compared to the hypothetical 2H-$WTe_2$ favors this configuration leading to a semimetallic ground state [5, 8]. The exact origin for the preference of the Td-structure instead of the 2H-polytype remains unclear. Earlier experiments on $WTe_2$ single crystals have revealed a metal-like behavior in the electrical resistivity below ~ 400 K beyond which the resistivity decreases slightly with temperature. The above together with temperature-dependent hall-coefficient and thermo-power measurements were earlier interpreted by Kabashima using a three-carrier semimetallic band model [9]. There is however no clear identification of these three bands owing to complex band structure with many interwoven bands as revealed by a study of Augustin *et al* based on angle resolved photoemission spectroscopy (ARPES) and density functional based augmented spherical wave calculations [8]. An extremely large unidirectional (along a-axis) positive magnetoresistance (MR) has been reported recently in single crystals of Td-$WTe_2$ [10]. MR in $WTe_2$ does not saturate even at very high applied magnetic fields and this is considered to be due to a perfect balanced electron-hole resonance in semimetallic $WTe_2$, as later complemented with high resolution ARPES study of low energy electronic structure [11]. The pronounced anisotropy in MR is ascribed to the uniaxial character of the Fermi surface and the proximity of balanced electron and hole Fermi pockets aligned along W-W chain direction in the k-space.

We have experimentally studied the electronic and phonon properties of bulk Td-WTe$_2$ as a function of temperature. One of the interesting properties of electronic transport is the metal-insulator switch-over in the electronic conductivity and a concomitant change in the sign of Seebeck coefficient from n-type to p-type at ~ 373 K. We have investigated the phonon properties of Td-WTe$_2$ using Raman spectroscopy and examined the effect of temperature and layer-thickness on the spectra. In order to understand the origin for stabilization of Td-polytype in WTe$_2$ and the experimentally observed features of electronic transport and phonon spectra, we have carried out first principle calculations, based on density functional theory.

**Experimental**

Synthesis and characterization: Polycrystalline WTe$_2$ was synthesized by annealing finely ground stochiometric amounts of W and Te elements (total weight of 4g) in an evacuated quartz tube at 650$^0$C for 12.5 hr and further at 1200$^0$C for 15 hr followed by cooling slowly to the ambient temperature. Powder X-ray diffraction patterns were recorded at various temperatures using Bruker D8 Advance diffractometer with Cu Kα radiation. Single crystals of Td-WTe$_2$ are synthesized by chemical vapor transport using the polycrytsalline WTe$_2$ powder.

Transport measurements: To study the temperature dependence of the transport properties, finely ground polycrystalline WTe$_2$ was compacted into bar and disc shaped samples under 30 MPa pressure. The samples were sintered at 500$^0$C for 10 hr in an evacuated quartz tube (10$^{-5}$ torr). The density of samples was determined by Archimedes method to be ~ 8.6 g/cm$^3$ i.e, about 92% of the expected theoretical density. Electrical conductivity (σ) and Seebeck coefficient (S) for the bar shaped sample were concurrently measured between room temperature (RT) and 673 K in Helium atmosphere by a four-probe method using ULVAC-RIKO ZEM3 instrument. Electrical resistivity with and without applied magnetic field were measured along the length of the bar from 300 K down to 3 K by a four-probe method using PPMS instrument. The applied magnetic field was perpendicular to the current direction in the bar. Magnetoresistance measurements were similarly carried out on single crystals of WTe$_2$. The thermal diffusivity (*D*) for the disc was measured between room temperature and 673 K using laser flash diffusivity method in Netzsch LFA-457 instrument and the total thermal conductivity (*κ*) was calculated using the formula $\kappa = C_p.D.\rho$ where *ρ* is the density of sample. The lattice contribution to the total thermal conductivity (K$_L$) was obtained using the relation, $\kappa = \kappa_{ele} + \kappa_{lat}$ where the electronic contribution $\kappa_e$ can be estimated by Wiedemann-Franz law: $\kappa_{ele} = \sigma LT$ where *L* is the Lorentz number. L is obtained from the reduced chemical potential (*η*) which is estimated by fitting Seebeck coefficient (*S*) as explained in the earlier reports [12, 13].

Atomic force microscopy (AFM): AFM of mechanically exfoliated few-layer WTe$_2$ flakes, deposited onto Si/SiO$_2$ (300 nm) substrates was carried out on using Bruker Innova Microscope instrument in tapping mode using antimony doped silicon tip with 10 nm resolution.

Raman spectroscopy: Raman spectroscopy was performed on few-layer WTe$_2$ flakes deposited on Si/SiO$_2$ and the disc-shaped compacted pellets in a temperature window ranging from 123 K to 400 K using LabRam HR microscope using Ar-laser (514.5 nm) excitation in back-scattering configuration. The laser power was adjusted to 2 mW by using a neutral density filter. Liquid N$_2$ cryostat was used for low-temperature measurements. N$_2$ gas was continuously purged throughout the measurement to prevent the condensation of moisture at low temperatures and possible oxidation at high temperatures.

**Computational details:**

Our first-principles calculations are based on density functional theory (DFT) as implemented in Quantum ESPRESSO package [14], in which the ionic and core-valence electron interactions are modeled with ultrasoft pseudopotentials [15]. The exchange-correlation energy of electrons is treated within a Generalized Gradient Approximation (GGA) with the functional parameterized by Perdew, Burke and Ernzerhof [16]. We use an energy cutoff of 35 Ry to truncate the plane wave basis used in representing Kohn-Sham wave functions, and energy cutoff of 280 Ry for the basis set to represent charge density. Structures are relaxed to minimize energy till the Hellman-Feynman forces on each atom are less than 0.02 eV/A. We use a periodic supercell to simulate a 2D sheet, including vacuum of 15 Å to separate the adjacent periodic images of the sheet. In self-consistent Kohn-Sham (KS) calculations of configurations of WTe$_2$ with monolayered form and bulk Td-structure unit cell, the Brilluoin zone (BZ) integrations are sampled over uniform meshes of 20x11x1 and 20x11x5 k-points respectively. Electronic structure is determined by including the spin-orbit interaction (SOI) through use of relativistic pseudopotentials using a second variational procedure [17]. Dynamical matrices and phonons at wave vectors on a 3x3x1 mesh in the BZ were determined using DFT linear response (Quantum ESPRESSO implementation [14] based on Green's function method). From these, dynamical matrices and phonons at arbitrary wave vectors in the BZ are obtained using Fourier interpolation.

**Results and Discussion**

A schematic of an orthorhombic unit cell of $WTe_2$ in figure 1(a) shows the vertical stacking of covalently bonded Te-W-Te triple layers along the c-axis via weak van der Waals interactions. A schematic of the structure viewed down the c-axis in figure 1(b) reveals the off-centering of W-atoms from their 'ideal' octahedral sites to form slightly buckled zigzag W-W chains running along the a-axis. Each W atom is surrounded by eight neighbors: six Te atoms and two W atoms. As seen from figure 1(a), the successive sandwich layers are sequentially rotated by $180^0$. The powder X-ray diffraction pattern of the as-synthesized polycrystalline $WTe_2$ confirms the formation of pure orthorhombic Td-$WTe_2$ phase with space group *Pnmn2$_1$* and the lattice parameters of a = 3.4770 Å, b = 6.2490 Å and c = 14.0180 Å.

We first note that the bulk Td structure of $WTe_2$ is (a) layered and (b) closely related to centrosymmetric 1T form (*c1T*) shown in figure 6(a)) [18]. We assess the structural stability of c1T monolayer and Td (bulk and monolayer) forms of $WTe_2$ through determination of their phonon spectra (figure 2). If a phonon spectrum exhibits phonon modes with imaginary frequencies ($\omega^2 < 0$), the structure is locally unstable (*i.e.*, it is not a local minimum, but a saddle point in the energy landscape); otherwise it is stable. The eigen displacements of the unstable modes precisely give the structural distortions that lower energy often lowering the symmetry. Our results for phonons of the stable structural forms are relevant and useful in Raman and infra-red (IR) characterization of these structures. Experimentally, bulk $WTe_2$ is known to adopt the Td structure and our calculated phonon spectrum of bulk $WTe_2$ exhibits no unstable modes, confirming its stability in the Td structure (figure 2(a)).

Group VI dichalcogenides with two non-bonding d-electrons usually adopt the trigonal prismatic coordination with filled $dz^2$ orbitals resulting in a semiconducting band gap. As mentioned above, Td-WTe$_2$ adopts an octahedral coordination around metal atoms. In the case of regular octahedral coordination (1T-structure), the t$_{2g}$ metal orbitals would be partially filled leading to metallic character. However, owing to the off-centering of W-atoms, the non-bonding t$_{2g}$ derived orbitals would experience some σ-bonding and split into bonding and antibonding orbitals to leave the density of states (DOS) minimum at the Fermi level resulting in a semimetal [8]. It is observed that the degree of distortion is commensurate with the splitting of the otherwise degenerate t$_{2g}$ derived orbitals [19]. ReS$_2$, a VII dichalcogenide, with three non-bonding d-electrons, is expected to be metallic both in trigonal prismatic (2H) and regular octahedral (1T) structures. However, ReS$_2$ exists in the 1T-structure with a semiconducting band gap of about 1.55 eV due to severe distortion of ReS$_6$ octahedra [20]. The distortion in Td-WTe$_2$ is lesser compared to 1T-ReS$_2$ rendering the former semimetallic.

Our analysis of the vibrational spectrum of the *c1T* polymorph reveals that WTe$_2$ is structurally unstable, exhibiting structural instabilities with imaginary frequencies of about 100i cm$^{-1}$ (figure 2b) at K and M points. The unstable mode at the high symmetry K-point of the BZ is doubly degenerate, while it is singly degenerate at the M-point of the BZ. We focus on the M-point instability and its consequences to the structure of the 1T form [18]. On distorting the *c1T* structure (figure. 6(a)) with eigen-displacements of its unstable mode at M-point, we get a √3x1 superstructure (see figure 6(c)) with zigzag chains of metal atoms. This distorted structure involving dimerization of metal atoms (where the M-M bonds are significantly contracted by 0.8 Å) is semimetallic (figure 6(d)). This distorted √3x1 superstructure is similar to monolayer of Td structure though the b/a ratio of experimental lattice parameters is 1.80 (as opposed to 1.73 here) due to coupling with strain. It is evident from phonon spectrum, that monolayered Td structure is locally stable. Weak instabilities near Γ point along the Γ-X and Γ-M directions involve long wavelength rippling of the 2-D planar structure (figure 2c), which is common to other 2-D materials [21, 22].

Bulk Td-WTe$_2$ has a periodic unit cell containing 12 atoms and belongs to the point group C$_{2v}$ and the space group *Pmn2$_1$*. There are 33 optically active modes at the Brillion zone center (at Γ-point) whose irreducible representations are:

$$\Gamma_{bulk} \rightarrow 11A_1 + 6A_2 + 5B_1 + 11B_2 \quad (1)$$

All the modes are Raman active because of the low crystal symmetry. According to the group theory, optical modes of A$_1$, B$_1$ and B$_2$ symmetry are Raman as well as IR active while the modes of A$_2$ symmetry are Raman active but IR inactive. The Raman spectrum of bulk Td-WTe$_2$ excited with a 514.4 nm laser shows five bands around 112, 118, 134, 165 and 212 cm$^{-1}$ (figure 3). The two prominent peaks are centered at 165 cm$^{-1}$ and 212 cm$^{-1}$. To assign the irreducible representations to calculated phonons at Γ, we obtained overlap (inner product) of a basis vector of an irreducible representation and eigen vectors of phonon modes (obtained using density functional perturbation theory). There are many modes with frequencies (tabulated in Table 1) close to those of the experimentally observed Raman bands, which make the assignment of the observed peaks nontrivial. For example, modes with frequencies of 162 cm$^{-1}$, 164 cm$^{-1}$ and 168 cm$^{-1}$ make the assignment of the observed intense peak at 165 cm$^{-1}$ subtle, *i.e.*, it can be A$_1$ or A$_2$ or B$_1$ mode. Similarly, modes with frequencies of 211 cm$^{-1}$ and 213 cm$^{-1}$ are close to the observed intense peak at 212 cm$^{-1}$ which therefore can be assigned to either B$_2$ or A$_1$ irreducible representation. The Raman tensor of these possible modes can facilitate the differentiation between these modes. The Raman tensor is calculated as a slope of the linear changes in electronic dielectric constant (the second derivative of the electronic density matrix with respect to a uniform electric field [23] with normal mode displacements. As bulk WTe$_2$ is semimetallic in nature (figure 5), its response to macroscopic electrical field is not finite or well defined and it is not straight forward to determine the Raman tensor directly. To this end, we have estimated the Raman tensor by constraining the occupation numbers of electrons so as to treat Td-WTe$_2$ as an insulator (i.e. number of occupied bands = number of electrons/2). We identify the modes around 168 cm$^{-1}$ and 207 cm$^{-1}$ as the ones with the largest Raman tensor, both belonging to A$_1$ symmetry. The observed intense Raman bands around 165 cm$^{-1}$ and 212 cm$^{-1}$ are in agreement with the theoretical estimates and hence labeled as the $A_1^{'}$ and $A_1^{''}$ modes respectively.

We have studied the effect of temperature on lattice vibrations in bulk polycrystalline Td-WTe$_2$. Figure 4(a) shows the Raman spectra of bulk Td-WTe$_2$ recorded at different temperatures. All the phonon modes are observed to soften with increasing temperature. Figure 4(b) shows the temperature-dependence of the intense bands at ~ 212 cm$^{-1}$ and ~ 165 cm$^{-1}$ obtained by fitting the bands with a lorentzian line shape. The data points were fitted using a Gruneisen model:

$$\omega(T) = \omega_0 + \alpha T$$

where $\omega_0$ is the peak position at 0 K and $\alpha$, the first order temperature coefficient obtained from the slope of the fit. The temperature coefficients ($\alpha$) for the intense peaks at ~ 212 cm$^{-1}$ and ~ 165 cm$^{-1}$ are -0.0081 and -0.0060 respectively. The observed phonon softening with increasing temperature is ascribed to the anharmonicity and decreased interlayer coupling at higher temperatures [24]. The higher value of $\alpha$ for the ~ 212 cm$^{-1}$ vibration relative to the ~ 165 cm$^{-1}$ vibration reflects higher sensitivity of the former to temperature and therefore, to the interlayer coupling.

Our calculations reveal that bulk and monolayer of Td-WTe$_2$ exhibit rather similar band structures (figure 5), both being semimetallic in nature. This behavior is in contrast to other TMDCs, in which a strong dependence of their electronic structure on the number of layers is seen [3]. From the partial density of states (DoS), we see that both the valence and conduction bands near Fermi energy levels (E$_F$) are composed of W-5d and Te-5p states indicating the covalency in W-Te bonding (figure 6(b,d)). Secondly, the spin-orbit coupling (SOC) included in calculations of electronic structure crucially influences even its qualitative features for example, the spin-split bands. It is clear from the crystal structure that dimerized chain of W atoms along a-axis (figure 1(b)), gives a semimetallic electronic structure along Γ-X direction in the BZ. Along M-Γ path, we find an indirect band gap close to Γ-point (see the zoomed picture in Figure 5). This indirect band gap increases from the bulk to monolayer by 0.1 eV.

Figure 7(a) shows the electrical conductivity (σ) of compacted polycrystalline WTe$_2$ measured between 3 K and 673 K. σ decreases from 1040 S/cm at ~ 3 K to 740 S/cm at ~ 373 K, exhibiting metallic behavior. Beyond 373 K, σ increases with temperature reaching 900 S/cm at ~ 673 K with semiconductor-like behavior [9, 10]. Figure 7(b) shows the temperature-dependence of Seebeck coefficient (S) measured between 300 K and 673 K. Interestingly, S varies from -9 µVK$^{-1}$ at 300 K to 35 µVK$^{-1}$ at 673 K with a change in its sign from negative (n-type) to positive (p-type) around the metal to insulator switch-over temperature of ~373 K. The metallic conductivity is due to the semimetallic nature of Td-WTe$_2$. The increase in conductivity above 373 K could arise from thermal excitation of carriers from the lower to the upper d-orbitals. As seen from figure 5(a), the distorted Td-WTe$_2$ structure has a narrow indirect band gap of around 0.05 eV near Γ point in the band structure seen along M-Γ path which allows thermal excitation of carriers leading to the observed increase in the electrical conductivity of Td-WTe$_2$ beyond 373 K. The change in the sign of S is consistent with the semimetallic nature [25-28]. The Fermi surface in Td-WTe$_2$ is experimentally observed to change drastically with temperature - the relative sizes of electron and hole pockets vary with temperature [11].

Figure 8 shows the total thermal conductivity (κ) measured between 300 K and 623 K as well as the extracted phonon ($κ_{lat}$) and electronic ($κ_{ele}$) contributions to the total κ (see experimental section). κ is nearly independent of temperature, ranging from 0.96 W/mK at RT to 1.06 W/mK at 623 K. The extracted values of $κ_{ele}$ and $κ_{lat}$ trend oppositely with the former increasing and the latter decreasing with temperature. Near 300 K, $κ_{lat}$ dominates $κ_{ele}$ and vice versa at higher temperatures. Preliminary magnetoresistance (MR) measurements showed a much smaller value of MR in the pellets of polycrystalline bulk sample relative to the single crystalline flakes of Td-WTe$_2$ due to the anisotropic nature of MR i.e, strong dependence of MR on the crystal-orientation [10].

Atomically thin layers of TMDCs, constituting a new class of graphene analogous 2D electronic materials, are promising next-generation functional materials featuring electronic, optical and mechanical properties of fundamental and practical interest and the applications ranging from spin- and valley-tronics to photocatalysis to flexible electronics [1-4]. Raman spectroscopy has been a powerful analytical tool for determining thickness and stacking of 2D layered materials [29], to study their thermal [24, 30] and mechanical properties [31] and to directly probe and monitor the charge-doping [32]. To understand the effect of layer-thickness on the lattice vibrations in Td-WTe$_2$, few-layer WTe$_2$ flakes were mechanically exfoliated (using a scotch tape) from bulk WTe$_2$ single crystals and deposited onto silicon substrates coated with 300 nm thick SiO$_2$ layer. The few-layer WTe$_2$ flakes were first identified using optical microscopy and the thicknesses of the flakes were precisely determined through atomic force microscopy (AFM). Figure 9 shows the optical images of exfoliated WTe$_2$ flakes deposited on Si/SiO$_2$ substrates and the corresponding AFM height profiles. A monolayer-thickness of ~ 0.7 nm was used for estimating the number of layers.

Figure 10(a) shows transmission electron microscope (TEM) image of mechanically exfoliated few-layer WTe$_2$ flake deposited on a carbon-coated copper grid. The high resolution TEM (HRTEM) image of a flake in figure 10(b) reveals an interlayer spacing of ~0.7 nm corresponding to the d-spacing of (002) planes. The HRTEM image of a flake with layers perpendicular to the electron beam is shown in figure 10(c). The fast Fourier transform (FFT) image of the region indicated in figure 10(c) projects a view down the [-1 0 1] zone axis with (111), (101) and (212) reflections (figure 10(d)).

Raman spectrum of 3-layer WTe$_2$ under 514.5 nm laser excitation reveals the six peaks centered at ~108, ~120, ~135, ~164, ~216 cm$^{-1}$. Figure 11(a) shows the Raman spectra recorded on exfoliated flakes of different thicknesses (3, 4, 5, 6, 10 and 25 layers) using 514.4 nm laser excitation. The intensity of all the Raman peaks increase with decreasing thickness. The shifts in various Raman peaks are plotted as a function of thickness (figure 11(b)). All the peaks are observed to soften with thickness except the one at ~165 cm$^{-1}$ which is nearly independent of thickness and the one at ~ 112 cm$^{-1}$ which stiffens with thickness. In 2H-MoS$_2$ and the related TMDCs, the out-of-plane A$_{1g}$ mode softens with decreasing thickness which is justified as being a consequence of decreasing interlayer van der Waals interactions and hence the effective restoring forces on atoms with decreasing thickness; the observed stiffening of the in-plane E$_{2g}$ mode with decreasing thickness is related to the influence of stacking on the intralayer bonding [33-36]. In our experiments, we find that the intense band at 212 cm$^{-1}$ is most sensitive to the number of layers, exhibiting an up-shift of about 4 cm$^{-1}$ in a 3-layer flake relative to bulk Td-WTe$_2$ whereas the intense band at 168 cm$^{-1}$ does not change with the number of layers though both belong to the same A$_1$ symmetry. To understand such dependence on thickness, we examine the displacements of W and Te in the eigenvectors of these modes (figure S2). Firstly, it can be seen that (i) the $A_1^{'}$ mode involves out-of-plane (z-direction) displacements of Te atoms and in-plane displacements of W atoms and (ii) the $A_1^{''}$ mode involves out-of-plane displacements of W atoms and in-plane displacements of Te atoms. Such mixing of in-plane and out-of-plane components of atomic displacements is a consequence of low crystal-symmetry in Td-WTe$_2$ and has been reported for a similar strongly distorted structure of 1T-ReS$_2$ [20]. Secondly, the Te atoms of the same plane vibrate in-phase (figure S2(a, b)) in the $A_1^{'}$ mode of vibration, while their motion is out-of-phase in the $A_1^{''}$ mode of vibration (figure S2(c, d)). Thus the $A_1^{'}$ mode seems to be more localized to a layer of WTe$_2$ and exhibits weaker or no dependence on the number of layers. To confirm the observed changes in Raman bands as a function of thickness, we compare the vibrational spectrum of monolayer Td-WTe$_2$ with that of the bulk Td-WTe$_2$. We find hardening of the $A_1^{''}$ mode by ~ 3 cm$^{-1}$ and softening of the $A_1^{'}$ mode by ~ 3 cm$^{-1}$ in the monolayer with respect to the bulk. While this is consistent with the observed trend in the $A_1^{''}$ mode, it suggests an additional compensating mechanism that governs the thickness-dependence of the $A_1^{'}$ vibrational mode, which needs further investigation. Furthermore, the ratio of the observed integrated intensities of $A_1^{''}$ to $A_1^{'}$ bands decreases from about 1.9 in bulk to about 1.1 in 3-layer flake reflecting the thickness-dependence of relative oscillator strengths of these modes.

Figure 12(a) shows the Raman spectra of 3-layer Td-WTe$_2$ recorded at different temperatures. All the Raman peaks are observed to soften with increasing temperature. Figure 12(b) shows the temperature-dependence of the observed intense peaks at ~164 cm$^{-1}$ and ~216 cm$^{-1}$ obtained by fitting the peaks with a lorentzian line shape. The temperature coefficients ($\alpha$) are -0.0098 and -0.0068 respectively for the intense peaks at 216 cm$^{-1}$ and 165 cm$^{-1}$ the former being higher than that in bulk WTe$_2$ and the latter being similar to that in bulk WTe$_2$.

## Conclusions

We have presented the electronic structure and vibrational properties of Td-WTe$_2$ in the monolayered and bulk forms, based on first-principles calculations. We bring out the connection of the Td structure with the 1T form of layered metal dichalcogenides and explain its stability in terms of the electronic and vibrational properties of the 1T form. From the electrical transport measurements, Td-WTe$_2$ is found to exhibit a metal to insulator switch-over in the electrical conductivity around 373 K, along with a change in the sign of the Seebeck coefficient from negative (n-type) to positive (p-type). The increase in conductivity beyond 373 K could arise from the thermal excitation of carriers across the indirect narrow band gap near the Γ point (seen along the M-Γ path). The Raman spectrum of bulk Td-WTe$_2$ shows five distinct bands around 112, 118, 134, 165 and 212 cm$^{-1}$. Based on symmetry analysis and calculated Raman tensors, we assign the intense bands at 165 cm$^{-1}$ and 212 cm$^{-1}$ to the $A_1^{'}$ and $A_1^{''}$ modes respectively. We have examined the effect of temperature and the number of layers on the Raman spectrum. A majority of the bands stiffen, and the ratio of integrated intensities of the $A_1^{''}$ to $A_1^{'}$ bands decreases with the decreasing number of layers while all the bands soften in both bulk and 3-layer samples with increasing temperature.

## Acknowledgments

M.K.J. acknowledges CSIR-UGC for fellowship. A.S. acknowledges JNCASR, India, for Fellowship. K.B. acknowledges DST Ramanujan fellowship, New Chemistry Unit and Sheikh Saqr Laboratory for funding. D.J.L. acknowledges the DST (India) for the Ramanujan Fellowship. U.V.W. acknowledges support from a J. C. Bose National Fellow-ship of DST (India).

**Figures and captions**

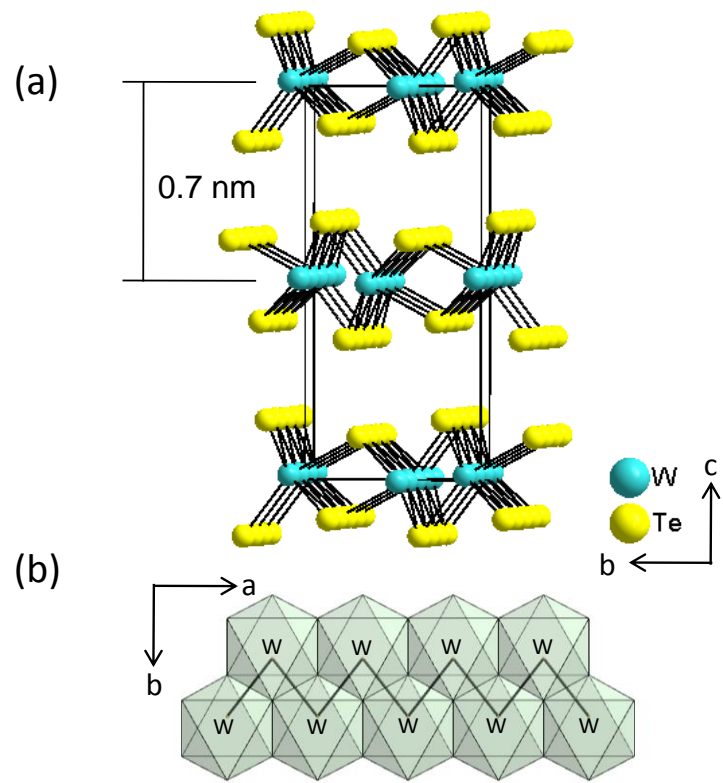

**Figure 1**. (a) Orthorhombic unit cell of Td-WTe$_2$ and (b) polyhedral representation of monolayer Td-WTe$_2$ showing W-W chains along a-axis.

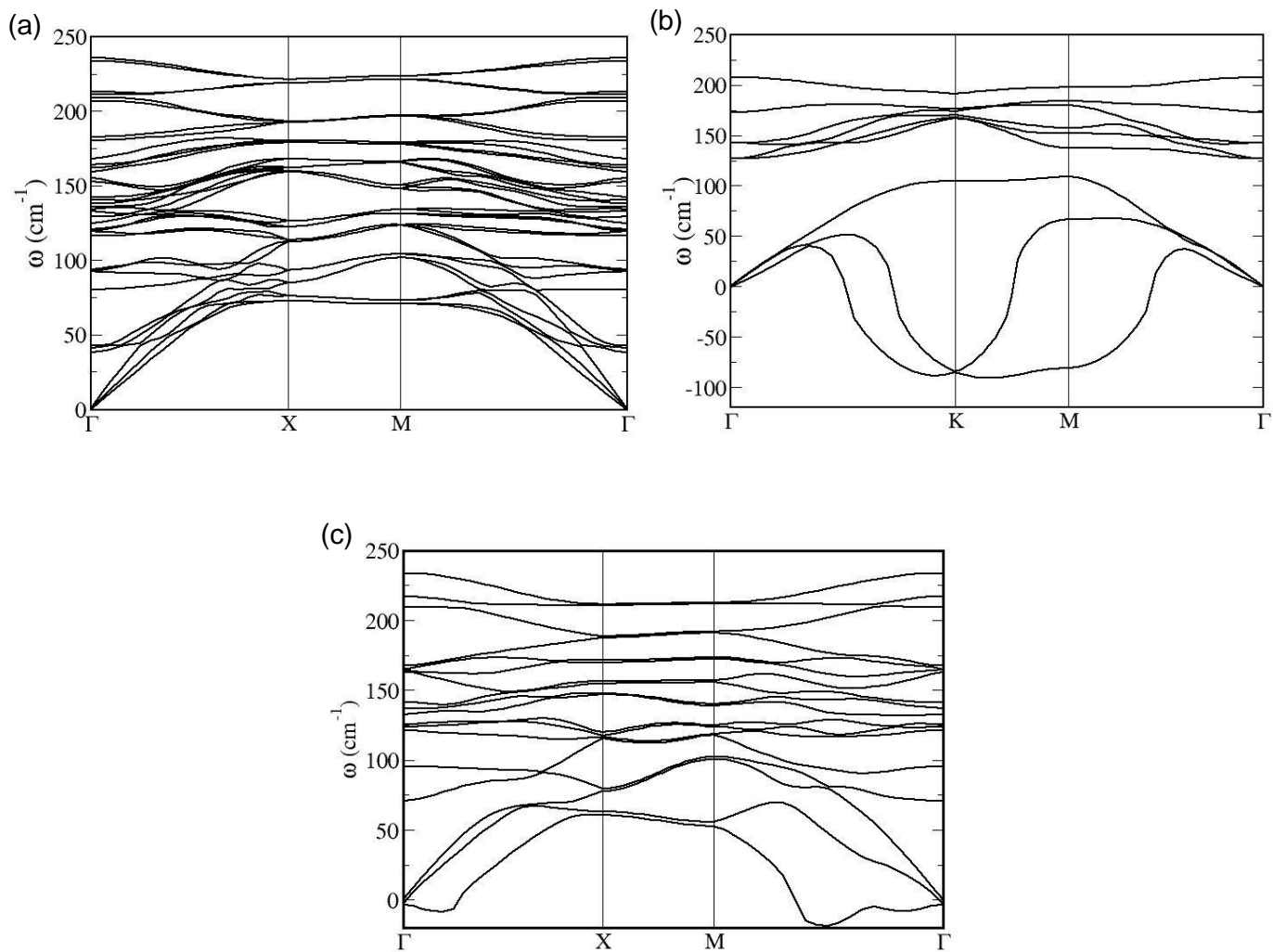

**Figure 2**. Phonon dispersion of (a) bulk Td-WTe$_2$ and monolayer WTe$_2$ with (b) c*1T* and (c) Td structures. Note that c*1T*- WTe$_2$ exhibits doubly degenerate and singly degenerate instabilities at K and M points respectively.

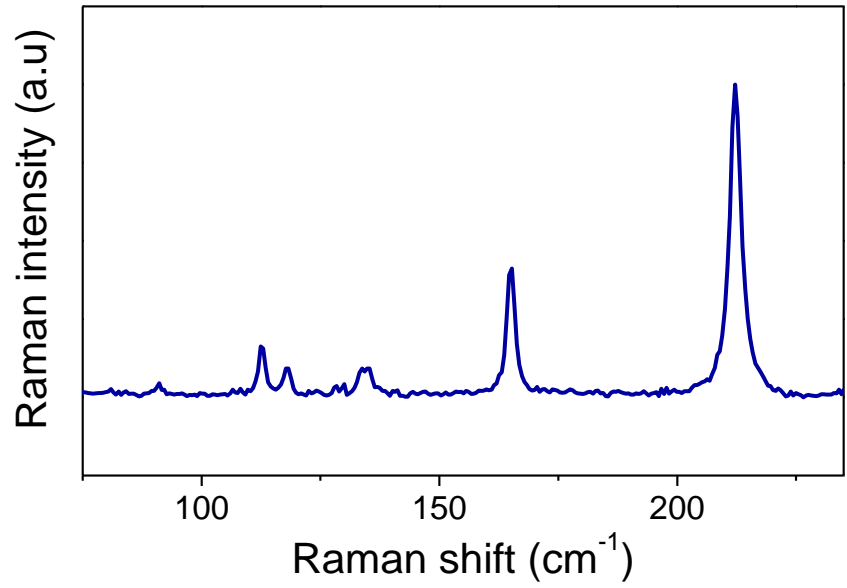

**Figure 3**. Raman spectrum of bulk polycrystalline Td-WTe$_2$ under 514.4 nm laser excitation

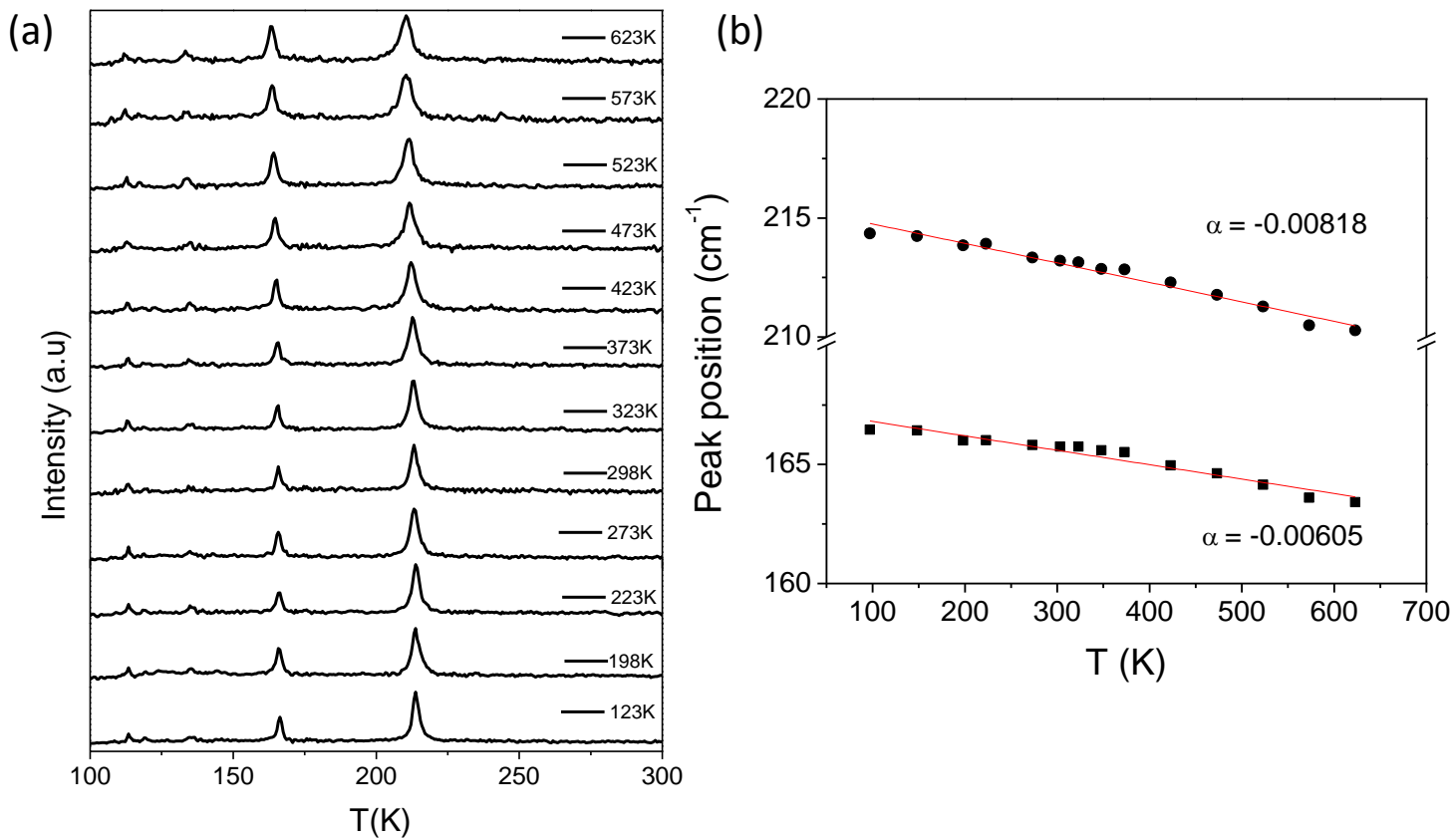

**Figure 4**. (a) Raman spectra of bulk polycrystalline Td-WTe$_2$ recorded at various temperatures using 514.4 nm laser excitation and (b) peak positions of the two most intense peaks as a function of temperature.

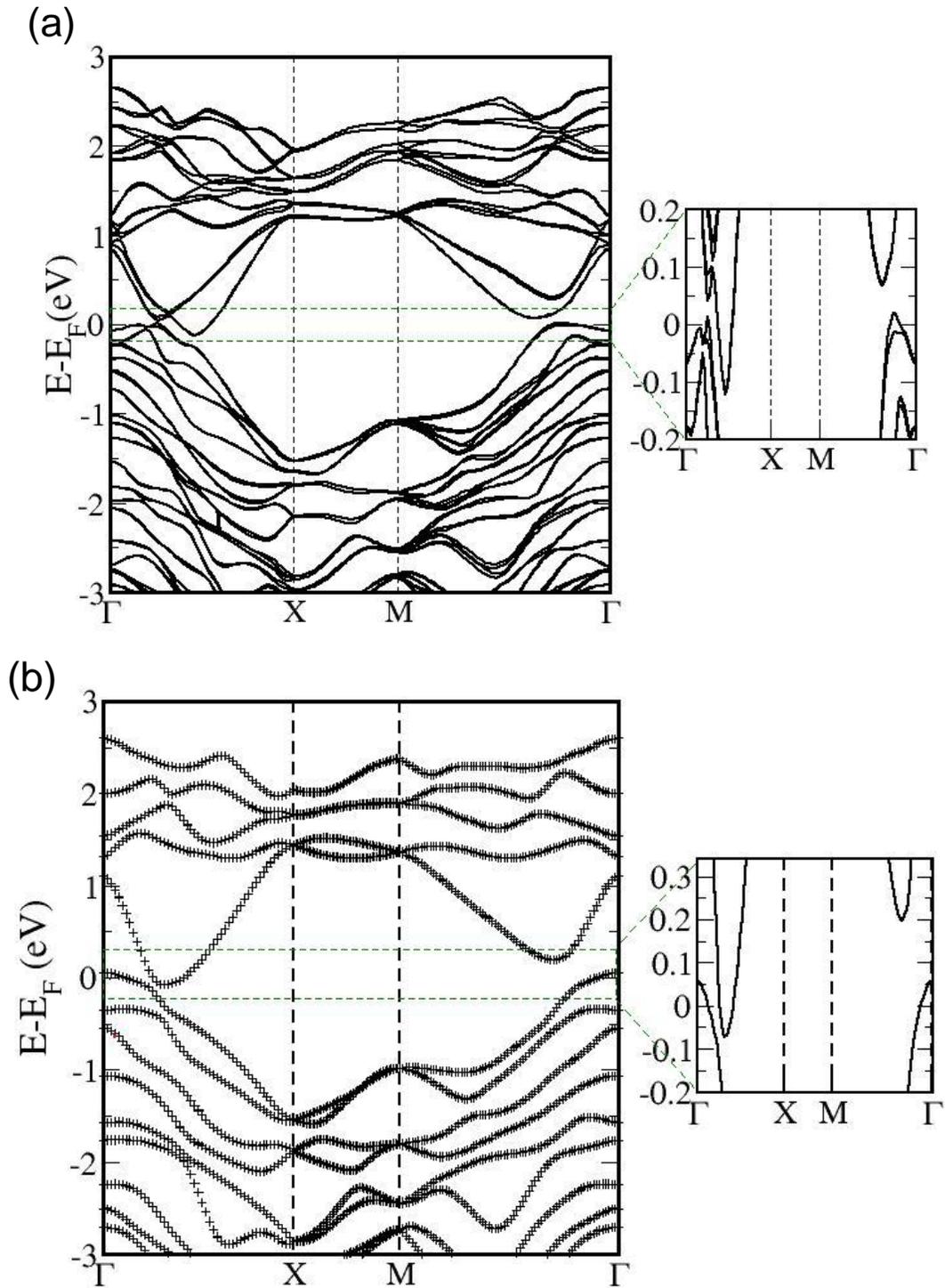

**Figure 5**. Electronic structure of Td-WTe$_2$ in bulk (a) and monolayer (b) forms. The region of band structures highlighted in green box is magnified on the right. Note that indirect band gap close to Γ point (along M-Γ path) in monolayer Td-WTe$_2$ is higher by 0.10 eV than that in bulk Td-WTe$_2$.
19

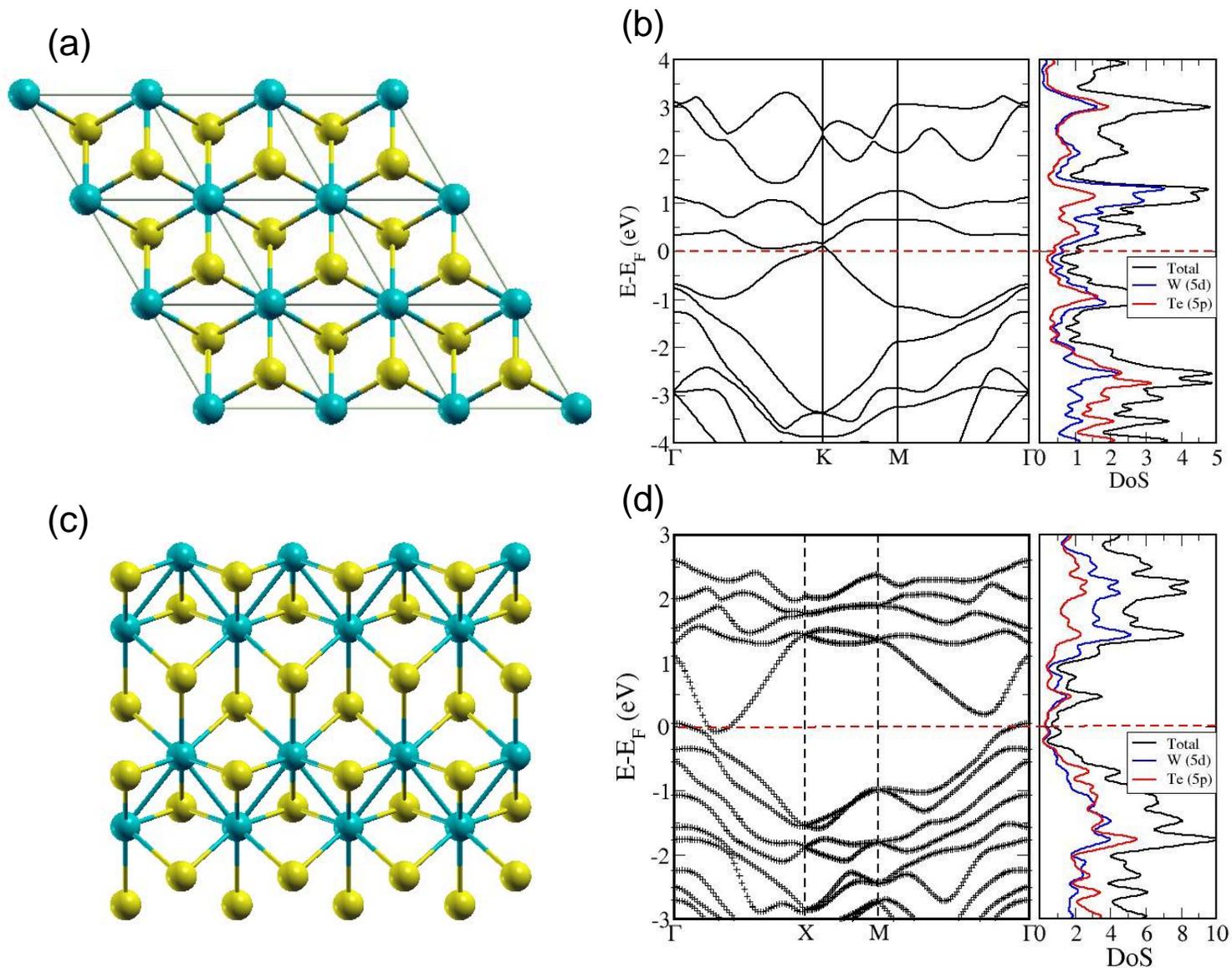

**Figure 6**. Crystal (top-view) and electronic structures of monolayer WTe$_2$: (a,b) c*1T* and (c,d) Td forms of WTe$_2$. c*1T* and Td structures of WTe$_2$ are metallic and semimetallic respectively. Note that spin-orbit coupling included in these calculations is crucial even for these qualitative properties of the electronic structure. W and Te atoms are shown in cyan and yellow spheres respectively.

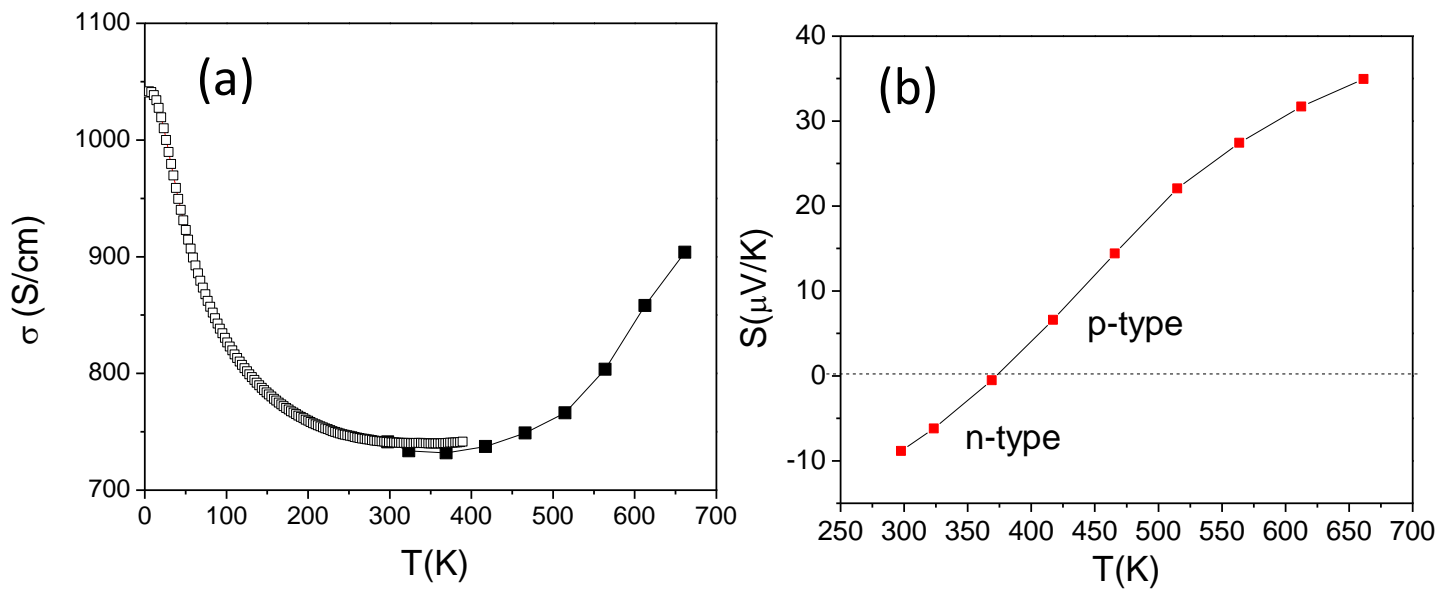

**Figure 7**. Temperature dependent (a) electrical conductivity ($\sigma$) and (b) Seebeck coefficient ($S$) of bulk polycrystalline Td-WTe$_2$.

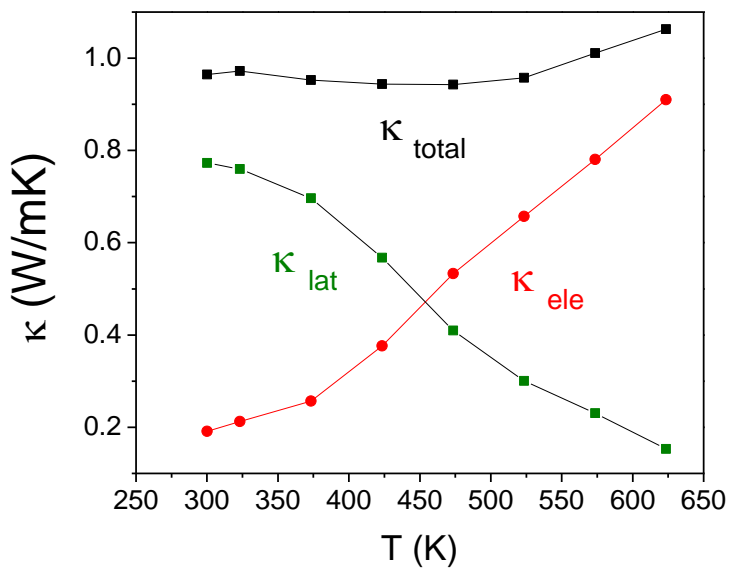

**Figure 8.** Total thermal conductivity ($\kappa$) along with individual electronic ($\kappa_{ele}$) and lattice contributions ($\kappa_{lat}$) of bulk polycrystalline Td-WTe$_2$.

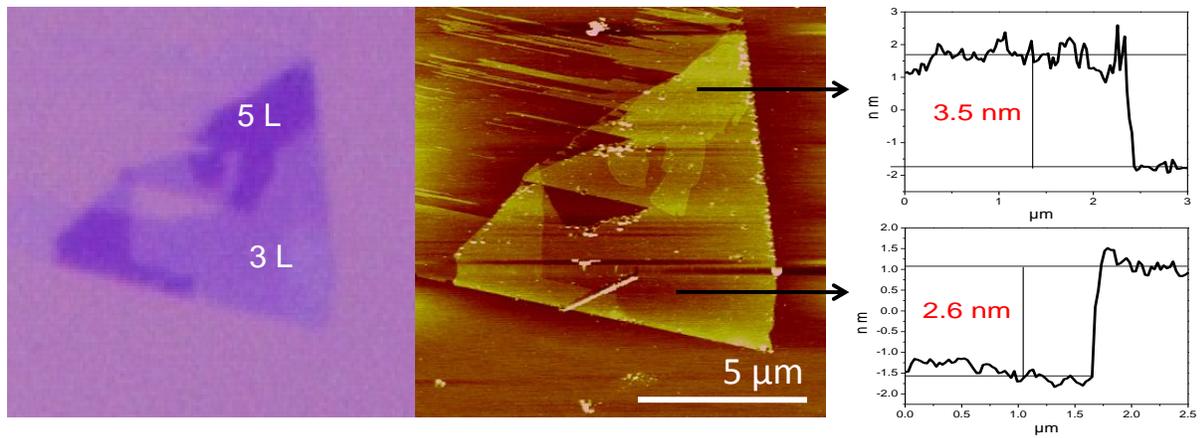

**Figure 9**. optical microscope images of few-layer Td-WTe$_2$ flake deposited on Si/SiO$_2$ substrates (left) and the corresponding AFM image (middle) and height profiles (right).

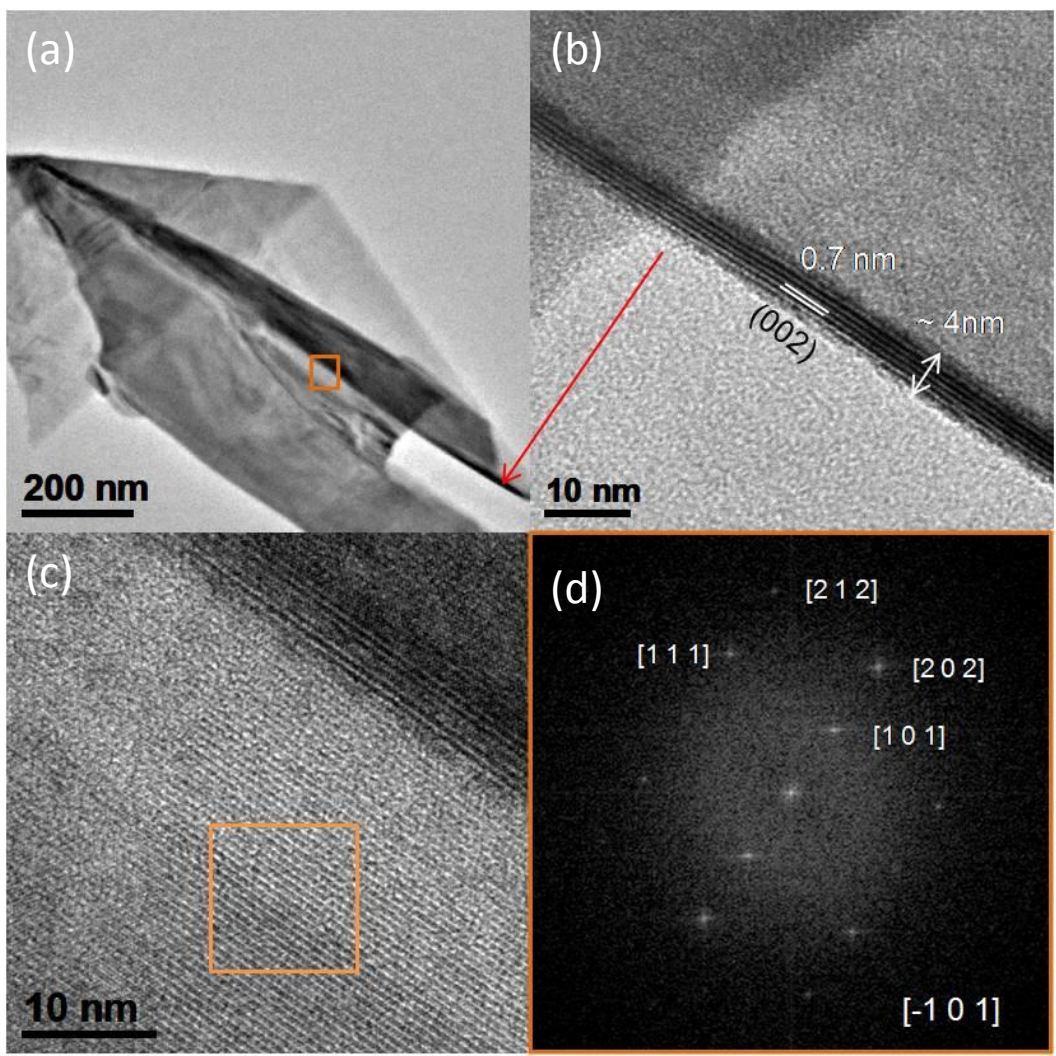

**Figure 10**. (a) TEM image of few-layer Td-WTe$_2$, (b) HRTEM image of region indicated by an arrow, revealing (002) planes with a d-spacing of ~ 0.7 nm, (c) HRTEM image of boxed region in (a) and (d) FFT image of the boxed region in (c).

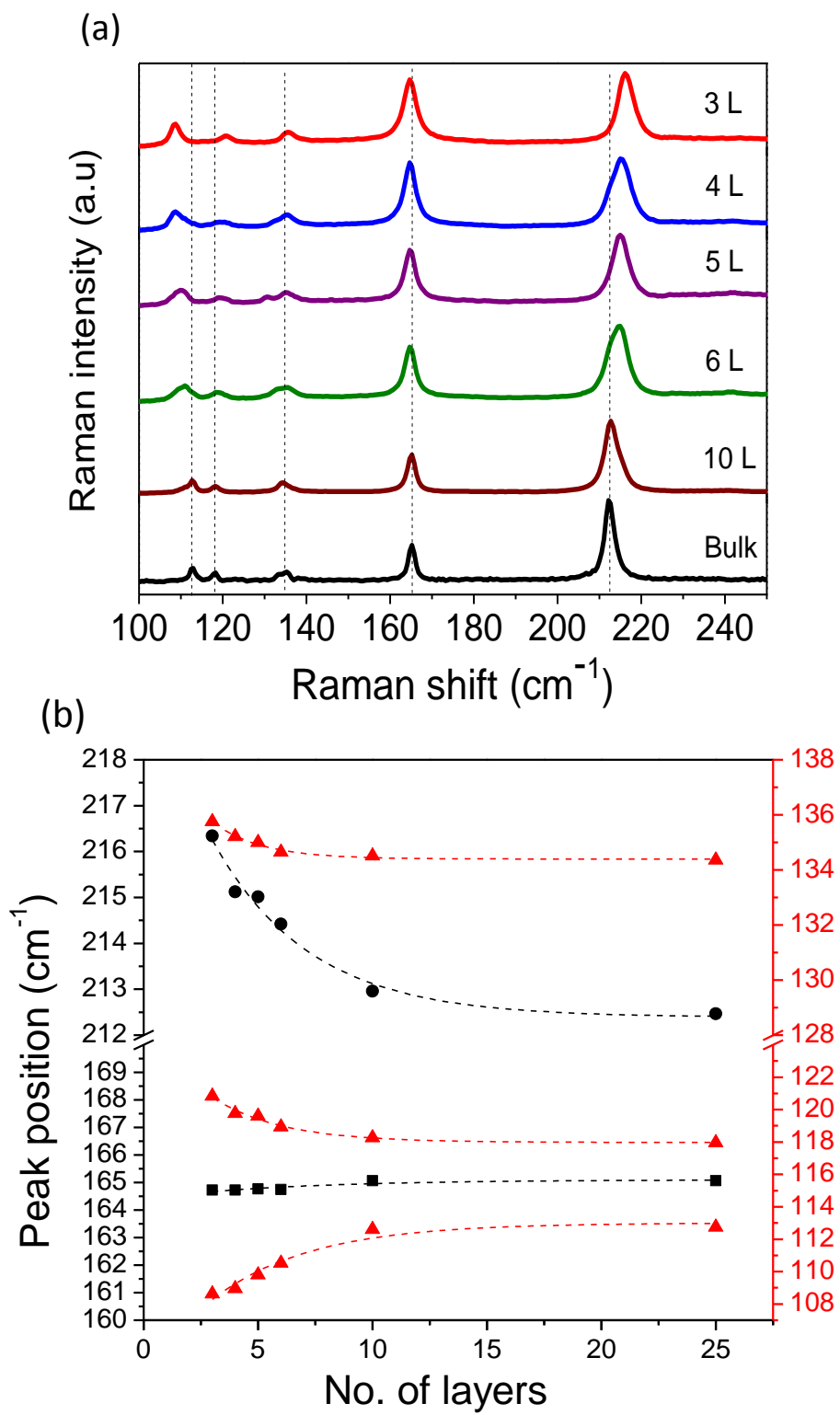

**Figure 11**. (a) Raman spectra of mechanically exfoliated Td-WTe$_2$ flakes as a function of no. of layers recorded using 514.4 nm laser excitation and (b) peak positions of various Raman peaks as a function of no. of layers. The dotted lines are guides to the eye.

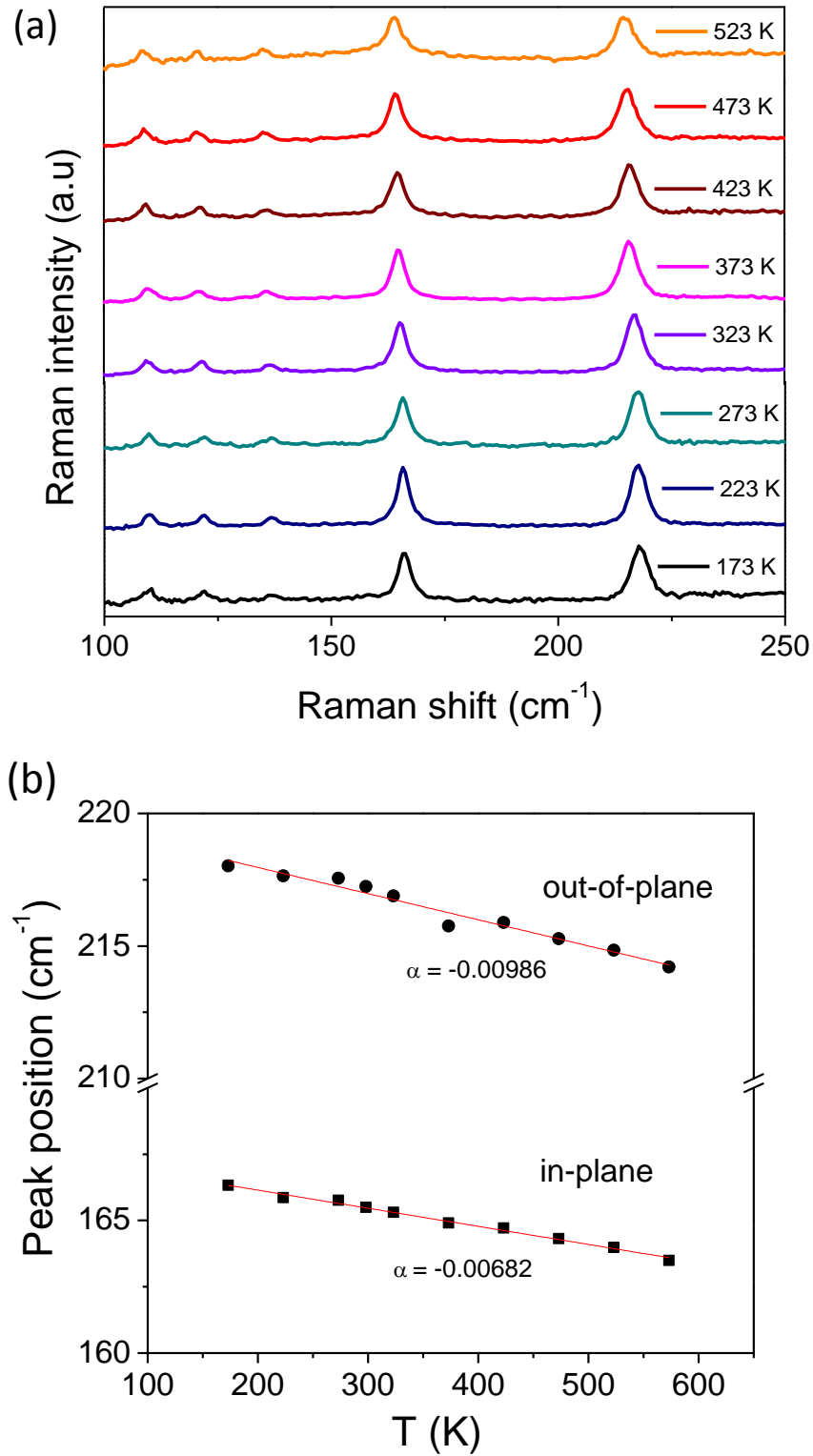

**Figure 12**. (a) Raman spectra of 3-layer Td-WTe$_2$ flake recorded at various temperatures using 514.4 nm laser excitation and (b) peak positions of the two most intense peaks as a function of temperature.

**Table 1**: Calculated phonon frequencies of Raman active modes of Td-WTe$_2$.

| A$_1$ Frequency (cm$^{-1}$) | A$_2$ Frequency (cm$^{-1}$) | B$_1$ Frequency (cm$^{-1}$) | B$_2$ Frequency (cm$^{-1}$) |
|---|---|---|---|
| 42 | 38 | 92 | 41 |
| 80 | 93 | 119 | 94 |
| 120 | 116 | 129 | 121 |
| 134 | 125 | 153 | 133 |
| 138 | 155 | 164 | 136 |
| 140 | 162 | | 142 |
| 168 | | | 159 |
| 183 | | | 180 |
| 207 | | | 209 |
| 211 | | | 213 |
| 233 | | | 236 |